\documentclass{article}

\usepackage{arxiv}

\usepackage[utf8]{inputenc} 
\usepackage[T1]{fontenc}    
\usepackage{hyperref}       
\usepackage{url}            
\usepackage{booktabs}       
\usepackage{amsfonts}       
\usepackage{nicefrac}       
\usepackage{microtype}      
\usepackage{lipsum}		
\usepackage{graphicx}
\usepackage{doi}
\usepackage{float} 
\usepackage{amsmath}
\usepackage{amssymb}
\usepackage{blindtext}
\usepackage{lineno}
\usepackage[numbers]{natbib}

\usepackage{float} 
\usepackage{booktabs}
\usepackage{caption}
\usepackage{subcaption}

\usepackage{newtxmath}
\usepackage{bm}
\usepackage{multirow}
\usepackage{color,soul}
\usepackage{ulem,xpatch}
\graphicspath{ {./images/} }
\usepackage{setspace}   
\usepackage{xcolor}

\title{CFD-DEM modeling of fracture initiation with polymer injection in granular media}

\author{ \href{https://orcid.org/0000-0001-8885-9511}{\includegraphics[scale=0.06]{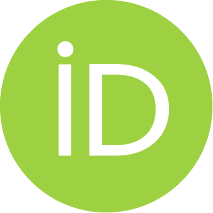}\hspace{1mm}Daniyar~Kazidenov }\\
	Department of Mathematics, \\
    School of Sciences and Humanities,\\
	Nazarbayev University\\
	Astana, Kazakhstan \\
	\texttt{daniyar.kazidenov@nu.edu.kz} \\
	\And
	\href{https://orcid.org/0000-0002-3958-8871}{\includegraphics[scale=0.06]{orcid.pdf}
    \hspace{1mm}Yerlan~Amanbek} \\
	Department of Mathematics,\\
    School of Sciences and Humanities, \\
	Nazarbayev University\\
	Astana, Kazakhstan \\
	\texttt{yerlan.amanbek@nu.edu.kz} \\
}

\renewcommand{\shorttitle}{\textit{arXiv} Template}

\hypersetup{
pdftitle={A template for the arxiv style},
pdfsubject={q-bio.NC, q-bio.QM},
pdfauthor={David S.~Hippocampus, Elias D.~Striatum},
pdfkeywords={First keyword, Second keyword, More},
}
\begin{document}
\maketitle

\begin{abstract}
 
We numerically study mechanisms and conditions of fracture initiation in granular material induced by non-Newtonian polymer solutions. A coupling approach of computational fluid dynamics and discrete element method is utilized to model the fluid flow in a porous medium. The flow behavior of polymer solutions and drag force acting on particles are calculated based on a power-law model. The adequacy of the numerical model is confirmed by comparing the results with a laboratory experiment. The numerical results are consistent with the experimental data presenting similar tendencies in dimensionless parameters that incorporate fluid flow rate, rheology, peak pressure, and confining stress. Results show that fluid flow rate, rheology, and solid material characteristics strongly influence fracture initiation behavior. Injecting a more viscous guar-based solution results in wider fractures induced by a grain displacement. A less viscous XG-based solution creates more linear fractures dominated by an infiltration. The peak pressure ratio between two fluids is higher in rigid material compared to softer material. Finally, the dimensionless parameters $1/\Pi_1$ and $\tau_2$, which consider fluid and solid material properties accordingly, are good indicators in determining fracture initiation induced by shear-thinning fluids. Our numerical results show that fracture initiation occurs above $1/\Pi_1 = 0.06$ and $\tau_2 = 2\cdot 10^{-7}$. 



\end{abstract}


\section{Introduction}


Injecting polymer solution into subsurface environments is a common process in many engineering fields such as enhanced oil recovery \citep{bera2020mechanistic}, hydraulic fracturing \citep{barati2014review}, groundwater and soil remediation \cite{luna2015pressure,omirbekov2023experimental}. 
When the pressure of injected fluid exceeds the strength of the rock, the resulting formation fracturing is a severe issue. Behavior of the fractures is complicated due to material characteristics and heterogeneity of the rock system \citep{kissinger2013hydraulic, shovkun2019propagation}. Moreover, the morphology of the fracture initiation and propagation is also related to stress conditions and permeability of the solid medium and fluid properties including flow rate and rheology \citep{warpinski1987influence, hurt2012parameters}.
Therefore, understanding fracture initiation mechanisms and the influence of fluid properties on fractures can help to predict further fracture propagation and optimize the best engineering design strategies.


Most experimental investigations of fracture initiation mainly focus on physical characteristics, shape, and macroscopic features of a fractured surface. For example, \citet{germanovich2012experimental} suggest that the peak pressure of injecting fluid and fracture aperture are primarily affected by confining stress of the formations. 
The ratio of peak pressure to confining stress at lower stresses is much greater than that of higher confining stresses. Fluid pressure is the dominant factor controlling fluid flow within a fracture. Therefore, fracture propagates perpendicular to the least confining stress. On the other hand, the permeability can be a more essential parameter than the confining stress in determining fracture behavior \cite{de2009fracture}. Experimental results indicate that fracture is inclined to propagate in less permeable zones showing infiltration in layers with higher permeability. \citet{bohloli2006experimental} observed that the fracture behavior of the sand is also strongly influenced by applied fluid additives and rheology. While injecting viscous Newtonian fluids and cross-linked gel resulted mainly infiltration, the cross-linked gel mixed with quartz powder and bentonite slurry produces clear fractures. 
\citet{huang2012granular} showed that the nature of fractures can be characterized by the fluid-grain displacement process due to various energy dissipation mechanisms. Therefore, the four distinct fracturing regimes such as a simple radial flow, infiltration-dominated, grain displacement-dominated and fingering regimes are classified according to characteristic times associated with fluid injection, hydro-mechanical coupling and viscoelastic behavior of the medium. 

In numerical analysis, the phase field method as a continuum approach is commonly used in recent years to study the fracture initiation and propagation problems \citep{almani2017multirate, wheeler2020ipacs}. \citet{lee2016pressure} developed a phase field model with adaptive mesh refinement to investigate pressure-driven and fluid-driven fracture propagation. A pressure diffraction equation coupled with a displacement–phase field, allows to examine complex fracture patterns including non-planar crack growth, joining and branching phenomena. \citet{brun2020iterative} proposed a novel phase field method with an iterative staggered scheme. The combination of an independent phase field function and elastic displacements of the solid material solves the problems associated with quasi-static brittle fracture propagation. Continuum approaches are generally limited to modeling fractures in granular materials where particle-level interactions predominate.
The discrete element method (DEM) coupled with computational fluid dynamics (CFD) becomes a popular instrument to study fracture initiation and propagation at microscopic level. While DEM clearly characterizes the reservoir sandstone as a granular medium consisting of individual particles, CFD accurately describes the behavior of fluid flow within the pores and fractures as well as its interaction with solid material. Many numerical studies focus on the influence of fluid properties on fracture initiation behavior. \citet{zhang2013coupled} observed that over certain ranges of fluid viscosity and injection rate the fluid flow demonstrates a transition between infiltration-dominated and infiltration-limited regimes, while the granular medium exhibits a transition between solid- and fluid-like behavior. Therefore, increase of fluid injection rate limits the infiltration resulting in localized failure and growth of fluid channels. Likewise, high-viscosity fluid dramatically increases the initiation pressure rate leading to a dominance of grain displacement rather than infiltration \cite{tomac2017coupled, sun2020fluid}.  Several numerical studies show that fracturing behavior is also significantly influenced by microscopic parameters of the medium and in situ stress conditions. Changes in the friction coefficient and Poisson's ratio slightly influence the fracture initiation. On the other hand, a solid material with an increased Young's modulus becomes more rigid, which is accompanied by a decrease in the width of the fractured zone \citep{li2020grain, sun2020fluid}. In situ stress conditions significantly influence the orientation of the fractures. In the case of isotropic stresses, the permeability of each pore may vary due to the heterogeneity of the local stress. Therefore, fractures may initiate along different paths depending on the magnitude of the stress. Under in situ stress anisotropy, fractures align along the direction of maximum principal stresses \citep{duan2018modeling}. Despite the advances in numerical studies of fracture initiation employing the CFD-DEM approach, most studies predominantly focus on using traditional Newtonian fluid models. Research related to a comprehensive understanding of the impact of non-Newtonian fluids on fracture initiation mechanisms has received insufficient attention. Current studies tend to neglect the influence of rheological parameters on the complex relationship between fluids and solid media. Addressing this issue is critical to further understanding fracture initiation under various fluid conditions and developing more accurate predictive models that incorporate non-Newtonian features.

In this work, we study the fracture initiation mechanisms and conditions by injecting non-Newtonian polymer solution into the granular medium. We utilize CFD-DEM framework, in which the power law model expresses the polymer flow behavior and corresponding drag force model calculates the interaction of particles with power-law fluid. First, the proposed numerical model is validated against the laboratory experimental data. The effect of fluid rheology, injection rate, applied stress conditions and solid material parameters on fracture initiation is then examined. Finally, we propose a fracture initiation criteria for shear-thinning fluids accounting for the fluid properties and micromechanical characteristics of the solid medium. 

This work is organized as follows: Section \ref{sec:2} presents governing equations of CFD-DEM model and numerical settings of the simulation. The validation part, obtained numerical simulation results and fracture initiation criteria based on dimensionless parameters are demonstrated in Section \ref{sec:3}. Finally, conclusions are summarized in Section \ref{sec:4}.

\section{Model formulation and numerical settings} \label{sec:2}

\subsection{Governing equations of CFD-DEM model}






In the numerical model, the fracture initiation simulations are performed by employing CFD-DEM coupling framework. While DEM calculates the particle motion and inter-particle interaction, CFD describes the behavior of fluid flow in the system. The relationship between the solid and fluid phases is established based on the particle-fluid interaction model. 

Particles in the system are accelerated due to collisions with each other, gravity, and interaction with the fluid. Newton's second law describes the particle motion which are expressed as follows: 
\begin{align}
m_i\frac{d \bm{v}_i}{dt} &= \bm{f}_{pf,i} +\sum_{j=1}^{k_c} \left( \bm{f}_{c,ij} + \bm{f}_{damp,ij} \right) +m_i \bm {g}\label{eq:1} \\
I_i\frac{d \bm{\omega}_i}{dt} &= \sum_{j=1}^{k_c} \bm{T}_{ij} \label{eq:2}
\end{align}

where $\bm{v}_i$, $\bm{\omega}_i$, $m_i$ and $I_i$ are translational and angular velocities, mass and inertia of a single particle, $\bm{f}_{pf,i}$ is the collection of forces exerted by fluid on a particle,  $\bm{f}_{c,ij}$, $\bm{f}_{damp,ij}$ and $\bm{T}_{ij}$ are the contact force, viscous damping force and torque acting between interacting particles, $m_i \bm g $ is the gravitational force, and $k_c$ indicates the number of particles that are in contact. 

We use modification of the JKR model to calculate the contact forces between particles, in which the cohesion (bonding) of particles is determined by surface energy density \citep{johnson1971surface}. Particle unbonding occurs when the applied stress exceeds the contact strength. As bonds are broken, no new bonds are formed. Detailed information of the model can be found in the following works \citep{rakhimzhanova2022numerical, kazidenov2023coarse}.


Apart from the the inter-particle interaction, the particles also come into contact with fluid. The particle-fluid interaction force  $\bm{f}_{pf,i}$ can be expressed as a total force exerted on a particle by fluid: 

\begin{equation}
\bm{f}_{pf, i} = \bm{f}_{d, i} + \bm{f}_{\nabla p, i} + \bm{f}_{\nabla \cdot \bm {\tau}, i} + \bm{f}_{Ar, i} \label{eq:2b}
\end{equation}

where $ \bm f_{d,i}$, $ \bm f_{\nabla p,i}$, $ \bm f_{\nabla \cdot \bm{\tau},i}$ and $ \bm f_{Ar, i}$ are the drag, pressure gradient, viscous and Archimedes forces, respectively.


The "model A" approach proposed by \citet{zhu2007discrete} is utilized to describe the fluid phase in the coupling model. Fluid flow behavior in a porous region ($\alpha_f$) is characterized by calculating locally averaged Navier-Stokes equations, in which the pressure gradient, viscous and gravitational forces are solely distributed between the solid and fluid phases:


\begin{equation}
 \left\{
\begin{array}{l}
\frac{\partial \alpha_f}{\partial t} + \nabla\cdot (\alpha_f \bm{u})=  0 \\ \vspace{0.01cm} \\
\frac{\partial (\rho_f \alpha_f \bm{u})}{\partial t} + \nabla \cdot \left( \rho_f \alpha_f \bm{u}\bm{u}\right)=  -\alpha_f \nabla p +\alpha_f \nabla \cdot \bm{\tau} + \rho_f \alpha_f \bm{g} + \bm{F}^{A}_{pf}  
 \label{eq:5} \\
 \end{array} 
  \right.
\end{equation}

where $\rho_f$, $\bm{u}$ and $p$ are the density, velocity and pressure of the fluid, respectively, $\alpha_f$ is the voidfraction of granular medium, $\bm{\tau}=\mu\left( (\nabla \bm{u})+(\nabla \bm{u})^T\right)$ is the stress tensor, where $\mu$ is the fluid dynamic viscosity and $\bm{F}_{pf}^{A}=\dfrac{1}{\Delta V}\sum_{i=1}^{n} \left( \bm f_{d,i} +  \bm{f}^{''}_i \right)$  is the volumetric particle-fluid interaction force, where $ \bm{f}^{''}_i$ is the sum of forces other than drag, pressure gradient and viscous forces, $\Delta V$ is the volume of a fluid cell.

The rheology of a polymer solution is expressed by a power-law model:

\begin{equation}
\mu (\dot{\gamma}) = K\dot{\gamma}^{(n-1)} \label{eq:power-law}
\end{equation}

where $K$ and $n$ are the consistency and flow behavior indexes, respectively.


The particle drag force is a critical component of particle-fluid interaction among other forces because it takes into account not only the voidfraction, particle size and shape, but also takes into account the rheology of the fluid.
In this study, we use a drag force model proposed by \citet{christopher1965power}, which describes the influence of power law fluids on multiple particles. 
\begin{equation}
\bm{f}_{d,i} = f \cdot  \frac{ \rho_f |\bm{u_i}-\bm{v_i}| (\bm{u_i} -\bm{v_i})(1-\alpha_f)V_{p,i}}{d_{p,i}\alpha_f} \label{eq:9}
\end{equation}
where $f$ is the Fanning friction factor,  $\bm{u_i}-\bm{v_i}$ is the fluid-particle relative velocity and $d_{p, i}$ is the particle diameter.
The Fanning friction factor in Equation (\ref{eq:9}) is expressed as follows:
\begin{equation}
f =  \frac{F_{CM} }{Re_{p, i}} \label{eq:10}
\end{equation}
where $F_{CM} = \frac{150}{12} \left(\displaystyle 9 +\frac{3}{n} \right)^{n}\alpha_f^{2(1-n)}(1-\alpha_f)^n$ and $Re_{p, i} = \frac{\rho_f d_{p, i}^n |\bm{u_i}-\bm{v_i}|^{2-n}}{K}$ is a particle Reynolds number, which considers the fluid rheology.

\subsection{Numerical settings} \label{sec:3_1}

This section describes the numerical settings employed for the model validation, modeling of the fracture initiation and  calculation of the permeability. While the validation and fracture initiation models are quite similar, the  calculation of the permeability uses different geometry and boundary conditions.   

\begin{figure}[H]
\begin{centering}
\includegraphics[width=1\columnwidth]{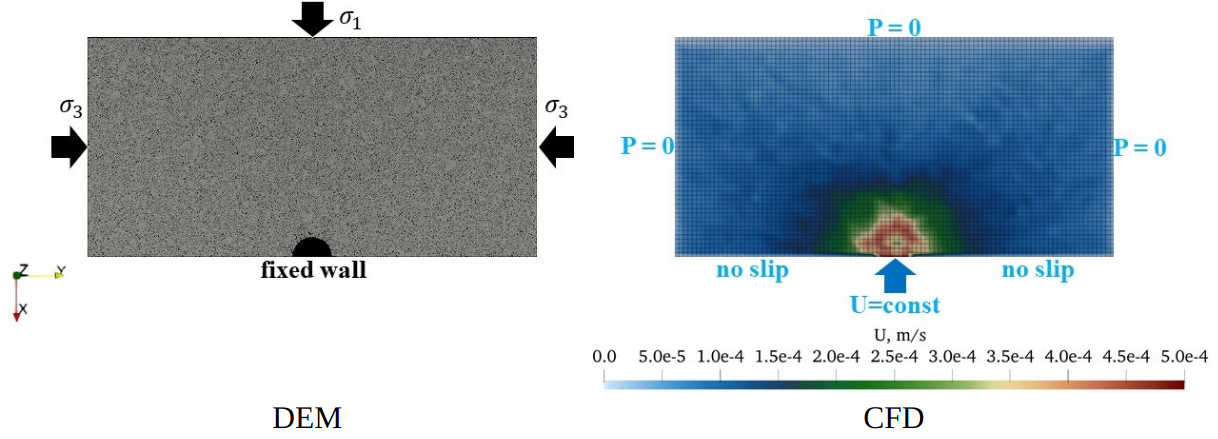}
\par\end{centering}
\caption{Numerical setup of the validation and fracture initiation models.\label{fig:sim_setup}}
\end{figure}

In the numerical model, we replicate the injection of the fluid into the Hele-Shaw cell. However, due to the higher computational cost of the simulations we use a single particle layer and half of the region. Despite the simplicity of the geometry, it was successfully validated in many applications and can capture basic mechanisms of fracture initiation \citep{li2023simulation, sun2018fluid}. The numerical setup of the validation and fracture initiation models is demonstrated in Figure \ref{fig:sim_setup}. It consist of separate DEM and CFD geometries, which are the same in size. The 0.002 m sized uniform particles are accommodated in single layer domain with dimensions $0.6 \times 0.3 \times 0.002$ m.The total number of particles equals to 45000. The material parameters of the simulated particles are provided in Table \ref{tab:mat_prameters}. The interaction between particles and wall is assumed as a frictionless. We apply $\sigma_3$ on both left and right sides and $\sigma_1 = 2\cdot\sigma_3$ on the back side of the domain. The fluid is injected at a constant rate into the domain through the wellbore in the negative x direction. The size of the wellbore is 0.05 m. We define the left, right and back sides as the outlets with corresponding P = 0 conditions. The initial conditions are $U (0) = 0$ and $P (0) = 0$. The solid-fluid phase coupling is utilized in an unresolved setting \citep{clarke2018investigation}, where a single fluid cell can contain multiple particles. The DEM time-step of the base case corresponds to the Rayleigh critical time-step \citep{li2005comparison} and equals to $\Delta t_{DEM} =  10^{-5}$ s. It decreases to  $10^{-6}$ s in cases where the Young's modulus of the material increases. The CFD model is configured with time-step values of  $10^{-4}$ and $10^{-5}$ s, respectively.  

\begin{table}[h!]
\fontsize{10}{12}\selectfont \caption{Material parameters of the granular medium used in the simulation.}
\begin{centering}
    \begin{tabular}{ p{5cm} p{2cm}}
\toprule 
Parameter \\
\midrule
 Density ($\mathrm {kg/m^{3}}$)      & 2650 \\
 Poisson's ratio   & 0.3   \\
 Young's modulus ($\mathrm {MPa}$)   & $1$   \\
 Friction coefficient  & 0.5   \\
 Surface energy density ($\mathrm {J/m^{2}} $)  & 1  \\

\bottomrule
\end{tabular}
  \label{tab:mat_prameters}
\par\end{centering}
\end{table}

\begin{figure}[H]
\begin{centering}
\includegraphics[width=1\columnwidth]{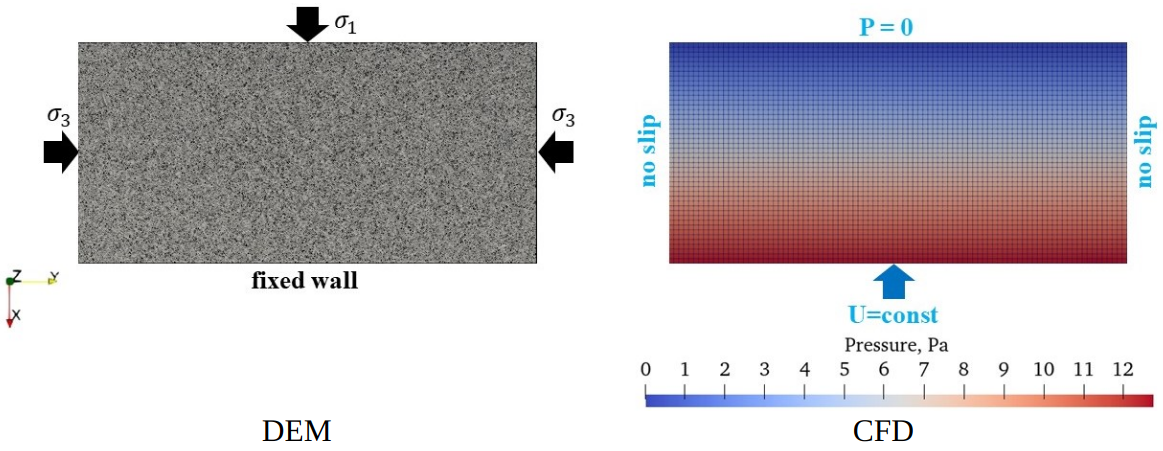}
\par\end{centering}
\caption{Permeability calculation simulation setup. \label{fig:permeability}}
\end{figure}

We perform additional simulations to calculate the permeability of the numerical granular medium for each corresponding confining stress and Young's modulus. The initial domain size is the same as in the main case. However, in the permeability calculation simulations, the wellbore is filled with particles and the fluid is injected from the whole plane at the bottom. The simulation setup is shown in Figure \ref{fig:permeability}. We use water as the injection fluid, and the particle drag force is calculated employing Di Felice model \citep{di1994voidage}. The fluid injection rate is $U = 10^{-3}$ m/s in all cases. We apply "no-slip" boundary condition to the left and right sides, and $P = 0 $ to the back plane, which is outlet.

Darcy Law is used to calculate the permeability ($k$):

\begin{equation}
k = \frac{Q \mu L}{A \Delta p}  \label{eq:12}
\end{equation}
where $Q$ is the flow rate, $\mu$ is the dynamic viscosity, $L$ is the length from the inlet to outlet, $A$ is the cross-section of the inlet and $\Delta p$ is the pressure difference between inlet and outlet. The calculated permeability values for different conditions are provided in Table \ref{tab:permeability}. 

\begin{table}[h!]
\fontsize{10}{12}\selectfont \caption{Calculated permeability values used in the simulation.}
\begin{centering}
    \begin{tabular}{p{3.4cm}p{1cm}p{1cm} p{1cm} p{1cm} p{1cm} p{1cm} p{1cm}}
\toprule 
 Young's modulus ($\mathrm {MPa}$) & \multicolumn{5}{c}{1} & 10 & 1000   \\
 Confining stress ($\mathrm {Pa}$) & 500 & 750 & 1000& 1250 & 1500 & $10^4$ & $10^6$ \\
 \midrule
 Permeability, $10^{-8}$ ($\mathrm {m^2}$) & 2.41 & 2.3 & 2.23 & 2.16 & 2.08 & 2.43 & 1.75   \\
\bottomrule
\end{tabular}
  \label{tab:permeability}
\par\end{centering}
\end{table}


\section{Numerical results} \label{sec:3}
\subsection{Validation of the model}

\begin{figure}[H]
     \centering
     \begin{subfigure}[b]{0.49\textwidth}
         \centering
         \includegraphics[width=1\textwidth]{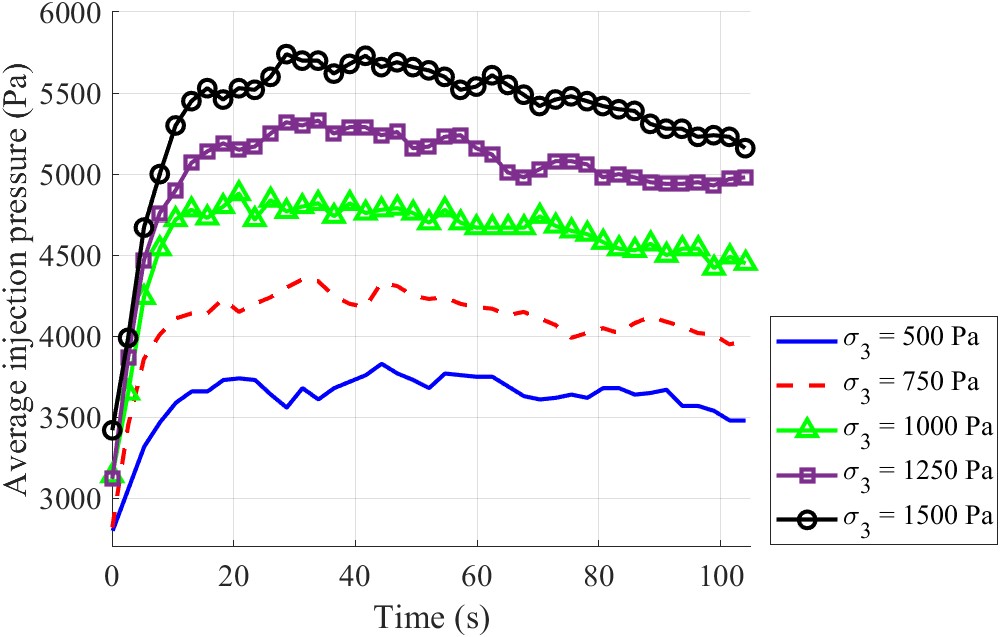}
         \caption{}
         \label{fig:peak_pressure}
     \end{subfigure}
     \hfill
     \begin{subfigure}[b]{0.49\textwidth}
         \centering
         \includegraphics[width=0.9\textwidth]{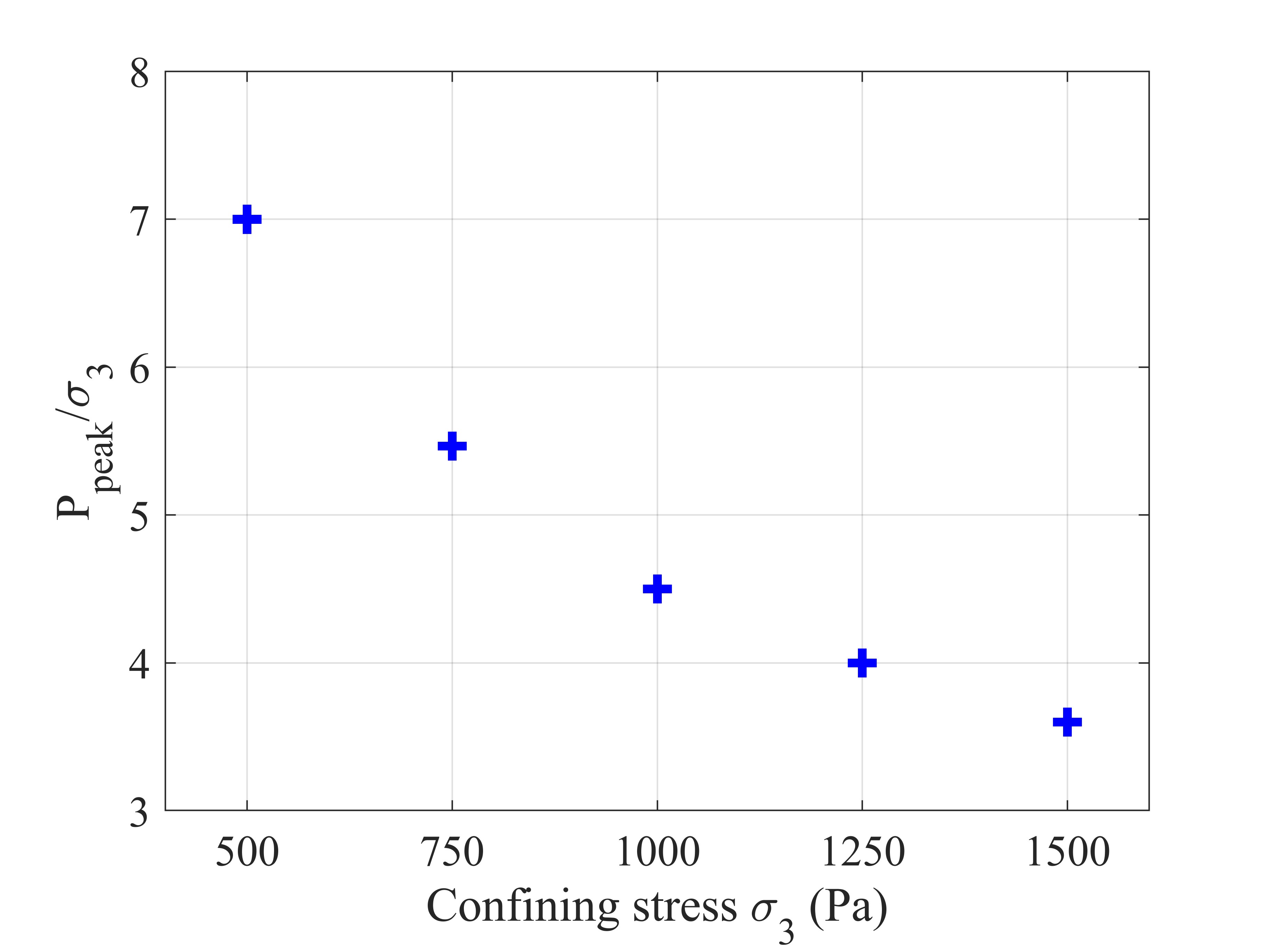}
         \caption{}
         \label{fig:ratio_of_peak_pressure}
     \end{subfigure}        \caption { (a) Time history of injection pressure (b) $P_{peak}/\sigma_3$ versus confining stress $\sigma_3$.}
        \label{fig:peak_pressures}
\end{figure}

\citet{germanovich2012experimental} showed that the peak pressure of the injected fluid and confining stresses are closely related to each other. The ratio of peak pressure to confining stress is extremely higher for low confining stresses and it decreases to about 3 at the higher stresses. To describe the fracture initiation for non-Newtonian fluids two distinct dimensionless variables based on peak pressure $p$ and confining stress $\sigma_3$ are proposed. Each variable includes other parameters such as flow rate $Q$, permeability $k$, and fluid rheological parameters (consistency index $K$ and flow behavior index $n$):
\begin{equation}
\Pi_1 = \frac{P_{peak}}{K} \left(\displaystyle \frac{k^{3/2}}{Q} \right)^n  \label{eq:peak}
\end{equation}
\begin{equation} 
\Pi_2 = \frac{\sigma_3}{K} \left(\displaystyle \frac{k^{3/2}}{Q} \right)^n  \label{eq:13}
\end{equation}
We validate the current numerical model against the experimental data provided by \citet{germanovich2012experimental}. We use guar-based polymer solution (Guar), which rheological parameters are provided based on power-law model ($K = 11$ Pa$\cdot$ s$^n$, $n =$ 0.41). For validation, the fluid is injected with the injection rate of $5\cdot10^{-4}$ m/s at various confining stresses $\sigma_3$  (500, 750, 1000, 1250 and 1500 Pa). We use the minimum injection rate at which fracture initiation occurs. The material Young's modulus is the same for all cases, which is $E = $ 1 MPa. Rest parameters and conditions are the same as provided in section \ref{sec:3_1}. \\
Figure \ref{fig:peak_pressure} shows the time history of the average injection pressure of a guar-based polymer solution at various confining stresses.   
The maximum value of injecting pressure (peak pressure) indicates the starting point of an extensive fracture initiation. The peak pressure value depends on various conditions such as flow rate, confining stress, fluid rheology, solid material characteristics, etc. We observe that increase in confining stress leads to higher peak pressures. At lower stresses, the fluid encounters less resistance to flow due to the higher permeability. On the other hand, under higher stress values, the fluid faces greater resistance and requires higher pressure to create fractures. The formation of fractures creates channels through which fluid flows easily, resulting in a decrease in injection pressure. While at lower stresses the pressure reduction tends to be gradual, at higher stresses the pressure drops sharply.  \\
Figure \ref{fig:ratio_of_peak_pressure} shows the ratio of peak pressure to confining stress $P_{peak}/\sigma_3$ for different values of applied confining stress. We observe that $P_{peak}/\sigma_3$ decreases with increase of confining stress, resulting in a value around 3.5 for higher stress. A similar trend is observed in the results of \cite{bohloli2006experimental, zhou2010experimental, germanovich2012experimental}.


\begin{figure}[H]
\begin{centering}
\includegraphics[width=0.6\columnwidth]{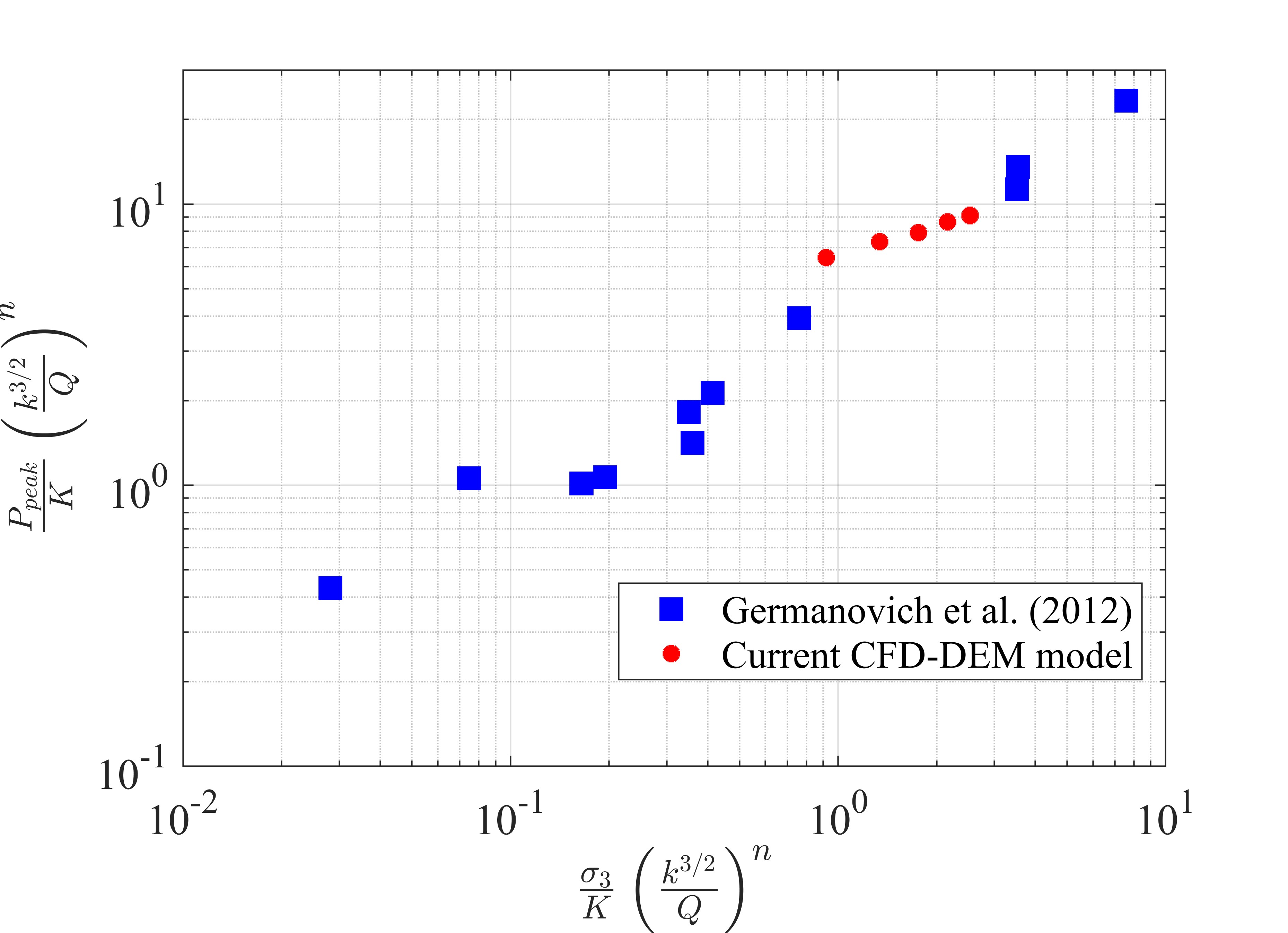}
\par\end{centering}
\caption{Comparison of dimensionless parameters between the experimental data and numerical results of guar-based polymer solution at $U = 5\cdot10^{-4} $ m/s and $\sigma_3$ = 500, 750, 1000, 1250 and 1500 Pa. \label{fig:dimensionless_guar}}
\end{figure}

Figure \ref{fig:dimensionless_guar} shows a comparison of the dimensionless peak pressure and confining stress values between the experimental data \citep{germanovich2012experimental} and the numerical results. The dimensionless variables are calculated using equations \ref{eq:12} and \ref{eq:13}. Despite the smaller scale, injection rate, and confining stresses, the numerical results are within the adequate range of dimensionless parameters calculated using the experimental data. These findings show that the current numerical model can be useful for the fracture initiation simulation induced by power-law fluids.



\subsection{Injection of different fluids} 

\begin{figure}[H]
\begin{centering}
\includegraphics[width=1\columnwidth]{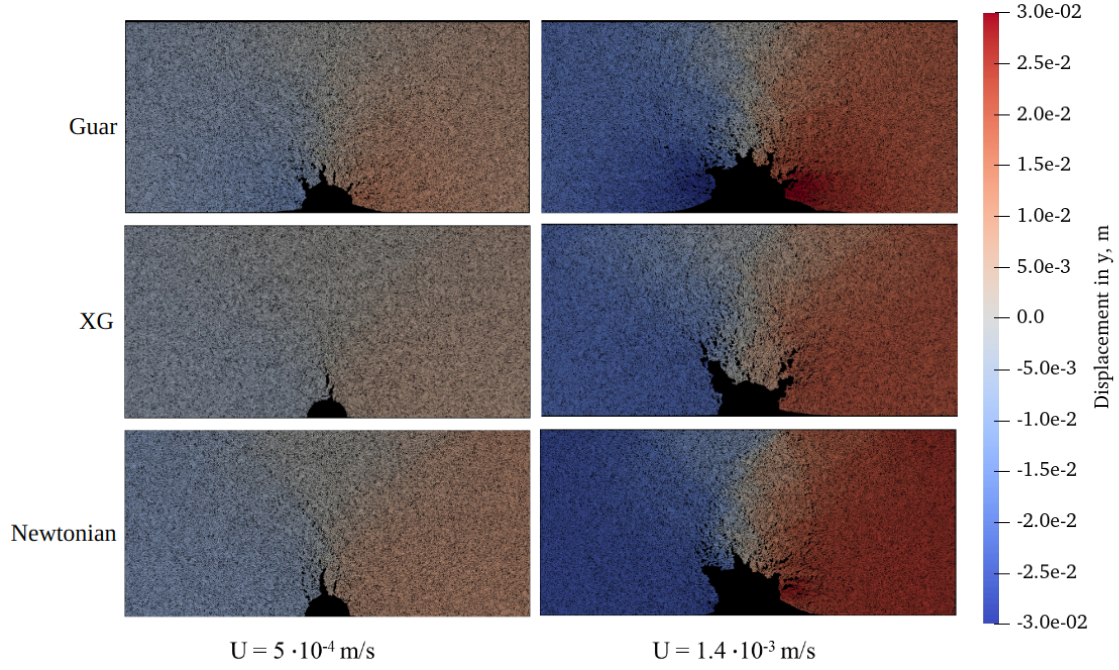}
\par\end{centering}
\caption{Comparison of fracture initiation by injecting different fluids. \label{fig:different_fluids}}
\end{figure}

To examine the behavior of fracturing at different fluids we compare fracture initiation of guar-based polymer with less viscous Xanthan gum (XG)-based polymer solution, which rheological parameters as follows: $K = 4.78$ Pa$\cdot$ s$^n$, $n$ = 0.1547 \citep{kazidenov2023polymersand} and non-Newtonian fluid with a constant viscosity of 3 Pa s. Figure \ref{fig:different_fluids} shows fracture initiation snapshots at 90 s by injecting different fluids at various injection rates. 
Depending on injecting velocity and various rheological parameters all three fluids have different fracture propagation shapes.
For example, at the lower injection rate ($5 \cdot 10^{-4}$ m/s), all fluids create small linear fractures oriented perpendicular to the lowest stress. Driving force of the fluids is sufficient to move particles to the less confined side. Guar and Newtonian fluids create multiple fractures due to higher viscosity, resulting in increased driving force. On the other hand, XG creates a single fracture due to lower viscosity. 
The higher injection rate results in a more significant particle displacement mechanism. For example, at injection rate of $1.4 \cdot 10^{-3}$ m/s, Guar and Newtonian fluids create wider fractures, characterized by the wellbore expansion with small cracks. Due to the significant driving force, fracture orientation is independent of stress, distributed radially along the wellbore. On the other hand, we observe a dependence on stress in fracture orientation induced by XG, which creates more linear fractures. 
The wellbore does not expand significantly compared to Guar and Newtonian fluid.



\begin{figure}[!htbp]
\begin{centering}
\includegraphics[width=1\columnwidth]{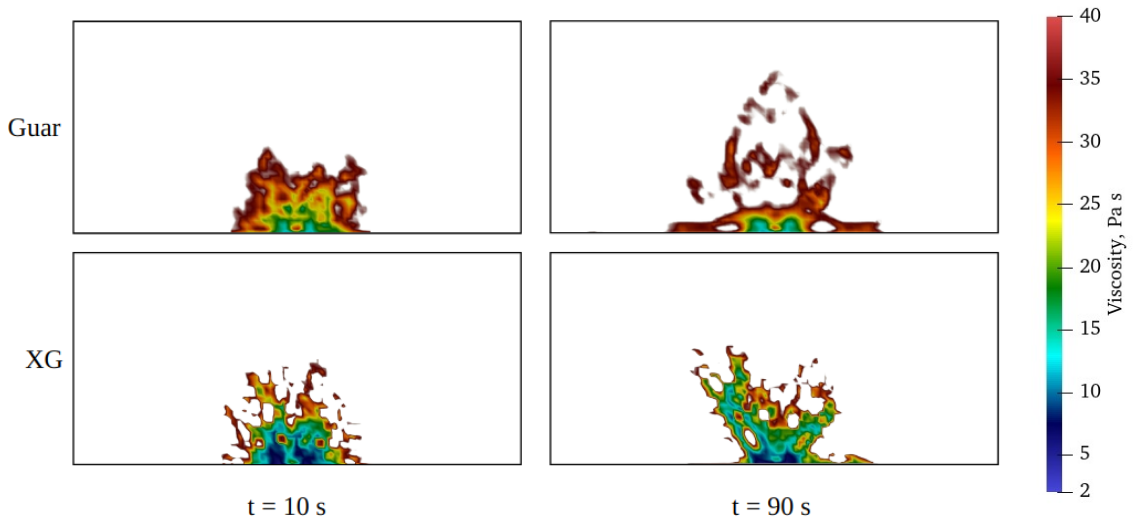}
\par\end{centering}
\caption{Viscosity of guar- and XG-based polymer solutions during fracture initiation at 10 s and 90 s. \label{fig:viscosity}}
\end{figure}

Figure \ref{fig:viscosity} demonstrates the viscosity changes of injected fluids at a rate of $1.4 \cdot 10^{-3}$ m/s. The snapshots emphasize only the critical place of the fracture initiation near the wellbore. The viscosity changes are shown at the starting moment of fracture initiation (10 s) and at the end of simulation (90 s). At 10 s, the injected XG has less viscosity (about 5 Pa s) than Guar (about 10 Pa s) within the wellbore. Accordingly, the fluid flow in a porous medium is also different for both. While XG flows with moderate viscosity (thicker green and cyan and thinner red areas), Guar flows with higher viscosity ranges (thicker red colored area). Due to less viscosity, XG infiltrates into wider zones, displacing the particles to the side with lower confining stress. Therefore, the fracture orientation has a linear pattern. On the other hand, Guar behaves in a grain displacement regime with less infiltration resulting in a wellbore expansion. At the end of simulation (90 s), both fluids flow within the created fractures with moderate viscosity. Similar to the moment of initiation, Guar flows with increased viscosity compared to XG at the fluid-particle interface.



\subsection{Influence of Young's modulus on fracture initiation} 
\begin{figure}[H]
\begin{centering}
\includegraphics[width=1\columnwidth]{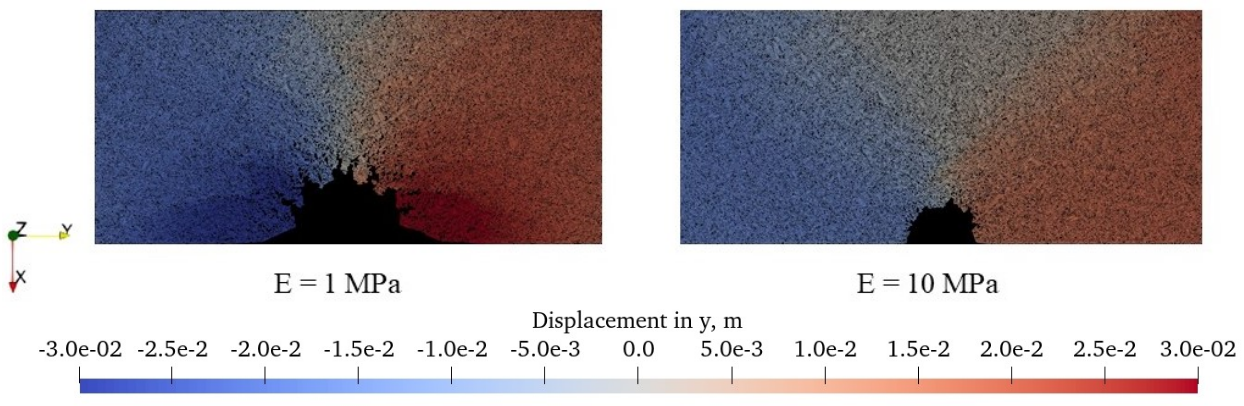}
\par\end{centering}
\caption{Fractures after 45 seconds of simulation at various Young's moduli by injecting a guar-based polymer solution.\label{fig:different_E}}
\end{figure}

In addition to fluid characteristics, the fracture aperture and orientation also depends on the micromechanical parameters of a granular medium, especially Young's modulus, which controls the deformation of particles. Figure \ref{fig:different_E} shows a comparison of fractures at particle Young's moduli ($E$) of 1 and 10 MPa created by a guar-based polymer solution. The injection rate of the fluid in both cases is the same, which is $3\cdot10^{-3}$ m/s. While at $E = 1$ MPa case the medium is compressed with $\sigma_3 = 1000$ Pa, at $E = 10$ MPa case, we increase $\sigma_3$ to 10000 Pa to maintain the same porosity and dimensions of the sample. As result, at 1 MPa case has a greater displacement of particles and wider propagation zone with fingering pattern. Several fractures occurred perpendicular to the wellbore, directed radially along the flow.  On the other hand, in the case of 10 MPa, we observe a minor expansion of the wellbore with the appearance of smaller cracks along it. These findings indicate that as Young's modulus increases, the particles become stiffer and thus the width of the fractured surface is limited. \\




\subsection{Peak pressures at different injection rates and Young's moduli} 
\begin{figure}[H]
     \centering
     \begin{subfigure}[b]{0.49\textwidth}
         \centering
         \includegraphics[width=1\textwidth]{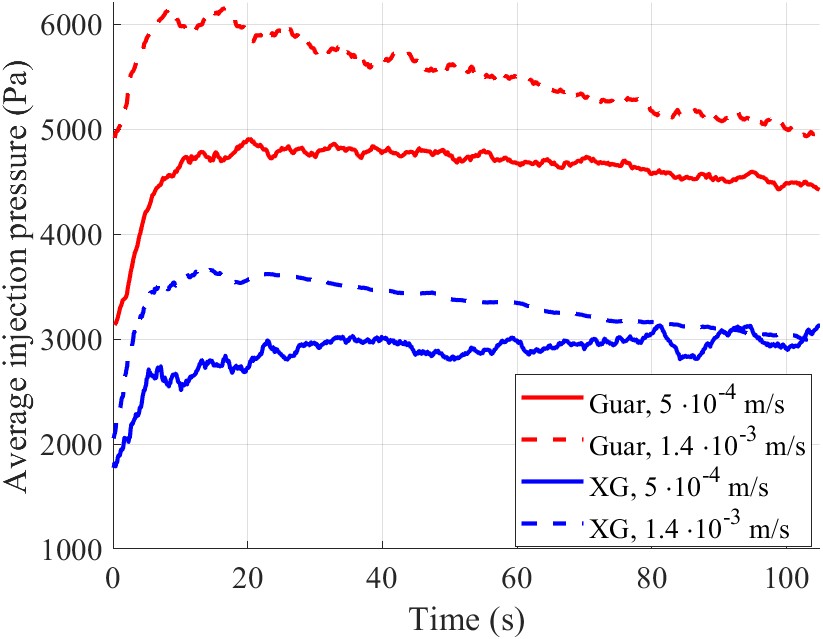}
         \caption{}
         \label{fig:DFRa}
     \end{subfigure}
     \hfill
     \begin{subfigure}[b]{0.49\textwidth}
         \centering
         \includegraphics[width=1\textwidth]{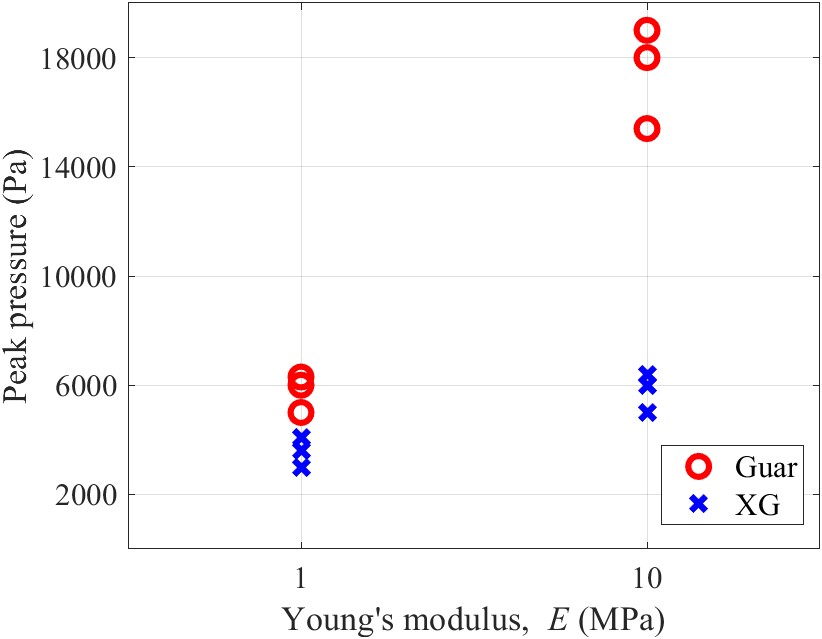}
         \caption{}
         \label{fig:E1_E10}
     \end{subfigure}
        \caption { (a) Injection pressure history at different injection rates. (b) Peak pressures at different Young's modulus for guar- and XG-based polymer solutions.}
        \label{fig:DFR}
\end{figure}

The injection rate has a greater influence on fracture initiation and therefore can be characterized by the injection pressure curve. Figure \ref{fig:DFRa} shows the change of the injection pressure for guar- and XG-based polymer solutions at different injection rates. The confining stress and Young's modulus for all cases are $\sigma_3 = 1000$ Pa and $E = 1$ MPa, respectively. In general, the injection pressure of guar-based solution varies at higher ranges compared to XG-based solution. At the lower injection rates, fracture initiation and expansion of the fracture surface occurs in slow manner. Therefore, lower injection rates are represented by late major peaks (around 5000 Pa in Guar and 3000 Pa in XG) and gradual decrease of injection pressure over time. On the other hand, at higher injection rates, the fracturing develops more dynamically resulting in earlier peak pressures (around 6100 Pa in Guar and 3600 Pa in XG). When the injection pressure reaches its peak, it decreases sharply since the fracture surface expands faster compared to lower injection rates. \\
Change of Young's modulus also influences the behavior of peak pressure. For example, at $E$ = 1 MPa, the peak pressure of Guar is about 1.66 times higher than that of XG, see Figure \ref{fig:E1_E10}.
However, when the material becomes more rigid ($E$ = 10 MPa), the ratio of peak pressures between two fluids increases to about 3.




\subsection{Fracture initiation criteria}  

\begin{figure}[H]
\begin{centering}
\includegraphics[width=0.9\columnwidth]{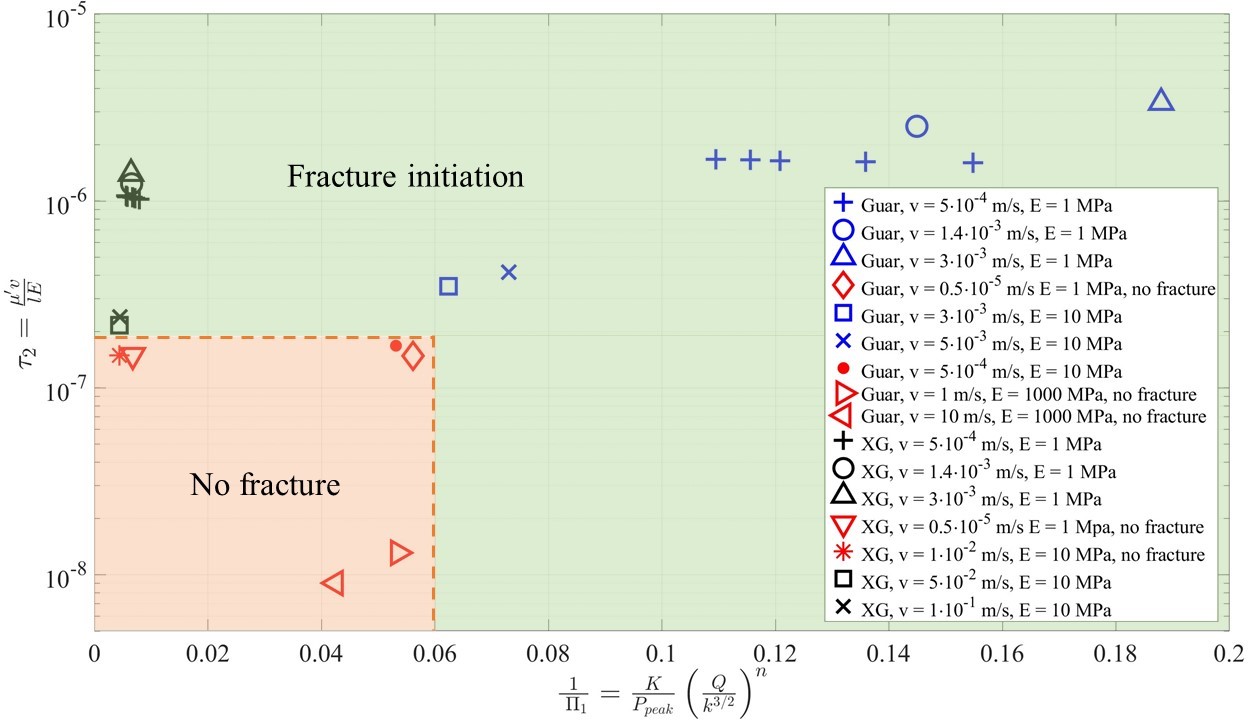}
\par\end{centering}
\caption{Fracture initiation criteria for guar- and XG-based polymer solutions based on dimensionless parameters $1/\Pi_1$ and $\tau_2$. Red symbols in orange region are no fracture cases. Blue and black symbols in green region are fracture initiation cases. \label{fig:fracture_model_guar}}
\end{figure}

From the investigations above, we observe that the fracture initiation behavior is influenced by many parameters such as fluid rheological characteristics, flow rate, peak pressure, confining stress and material Young's modulus. The fracture initiation regime in unconsolidated granular media can be described by using dimensionless parameters which consist of those parameters. \citet{huang2012granular} proposed dimensionless time $\tau_2$, which characterizes the viscoelastic behavior of the medium:


\begin{equation}
\tau_2 = \frac{\mu' v}{lE} \label{eq:tau_2}
\end{equation}

where $\mu' = \mu f(\phi)$ is the apparent viscosity of a fluid and grain mixture correlated with fluid viscosity $\mu$ and function of grain volume fraction $f(\phi)$, $v$ is the injection velocity, $l$ is the characteristics length of the inlet and $E$ is the solid material Young's modulus. At the same flow rate, when $E$ increases the $\tau_2$ decreases, and the medium behaves as solid, which leads to a regime with a predominance of infiltration. When $\tau_2$ increases by the decrease of $E$, the medium is more deformable resulting in a grain displacement-dominated regime. We define $\mu$ as the viscosity value at a given shear rate. The fracture initiation occurs due to the acting drag force on the particles by fluid with maximum velocity at the inlet, where the shear rate is maximum as well. Therefore, we measure $\mu$ at the maximum velocity, which corresponds to the inlet of the domain. 
We consider $f(\phi) = \phi_i$, where $\phi_i$ is the initial grain volume fraction before injection. For the second parameter we select the inverse of the previously introduced dimensionless parameter ($1/\Pi_1$) in Eq. \ref{eq:peak}. Both parameters vary in complex manner. While $\tau_2$ mainly determines the micromechanical characteristics of a granular medium, the behavior of $1/\Pi_1$ depends on fluid properties including flow rate, injection pressure and rheology. 
\\
Figure \ref{fig:fracture_model_guar} shows a fracture initiation criteria for guar- and XG-based polymer solutions defined by a combination of $1/\Pi_1$ and $\tau_2$ dimensionless parameters. The case of $5\cdot10^{-4}$ m/s at $E = 1$ MPa is presented with different confining stress ($\sigma_3$) values ($500, 750, 1000, 1250$ and $1500$ Pa). The remaining cases are presented with only a single stress value.  
The cluster of cases closer to the origin (red color symbols), which do not create fractures, forms the fracture initiation threshold outlined by the orange box. The threshold values are established at 0.06 and $2\cdot 10^{-7}$ for  $1/\Pi_1$ and $\tau_ 2$, respectively. Cases above the threshold (blue and black colors) are more likely to yield in fractures. Increase of flow rate and decrease of Young's modulus distances the initiation cases from the threshold. At the same injection rates, $\tau_ 2$ for both fluids has approximately the similar range of values, especially at Young's modulus $E = 1$. On the other hand, the values of $1/\Pi_1$ vary significantly between both fluids since they are based mainly on the fluid properties. Particularly, the $1/\Pi_1$ values of the guar-based solution are widely scattered  due to the greater magnitude of $n$ (0.45) compared to XG. In XG, a smaller value of $n$ (0.15) diminishes the effect of other parameters localizing the $1/\Pi_1$  values at closer distances.  
Both fluids do not create fractures at maximum value of Young's modulus ($E = 1000$ MPa), which is a replication of a real sandstone sample from a Kazakhstan reservoir \citep{kazidenov2023coarse, rakhimzhanova2019numerical, shabdirova2016sample}. At lower Young's modulus ($E = 1$ MPa), the fracture initiation occurs at similar injection rates for both fluids. As the $E$ increases to 10 MPa, the XG-based solution requires more than ten times higher injection rate than Guar to create fractures. The change of grain volume fraction at the same $E$ has significant impact on $1/\Pi_1$ compared to $\tau_2$, notably in a guar-based solution. We observe a dramatic movement to the right of $1/\Pi_1$ values when the stress is decreased. At the same conditions $\tau_2$ values change slightly. In particular, while fracture initiation induced by Guar-based solution is influenced equally by both dimensionless parameters, in XG, the effect of $\tau_2$ on fracture initiation is more significant than $1/\Pi_1$. These findings show that the fracture initiation depends on many factors associated with fluid characteristics as well as granular media. 



\section{Conclusions} \label{sec:4}

The primary aim of this work is to numerically investigate the basic mechanisms and conditions of fracture initiation in granular media by injection of shear-thinning polymer solutions. The fluid flow and drag force acting on particles are calculated considering rheological parameters of power-law fluids. The numerical model is initially validated against the laboratory experiment. Obtained results are in good agreement with experimental data showing similar behavior of dimensionless parameters, that account for fluid flow rate, rheology, peak pressure and confining stress. Moreover, similarly to the experiment, the ratio of peak pressure to confining stress increases at lower stresses but decreases at higher stresses. \\
Injection of different fluids demonstrate that the fracturing mechanisms are influenced by the fluid flow rate and rheology. At lower flow rates, all fluids create small linear fractures. At higher flow rates, while a more viscous guar-based solution creates wider fractures dominated by a grain displacement, the orientation of fractures induced by a less viscous XG  behaves more linearly dominated by an infiltration. Results show that the deformation characteristics of the solid material also impact the fracture aperture and orientation. The surface of fractures in a soft granular medium is wider with a radially distributed fingering pattern along the wellbore. On the other hand, a more rigid material reduces the width of fractured surface. The injection pressure during fracture initiation is higher for Guar compared to XG. Moreover, the peak pressure ratio between the fluids increases when fracturing occurs in a more rigid material. \\
Finally, we propose a fracture initiation criteria for shear-thinning fluids based on dimensionless parameters $1/\Pi_1$ and $\tau_2$, which take into account fluid properties and micromechanical characteristics of the granular medium, respectively. Fracture initiation occurs above the threshold values of $1/\Pi_1 = 0.06$ and $\tau_2 = 2\cdot 10^{-7}$ in both fluids. We observe that in a less viscous fluid $\tau_2$ has a much greater impact on fracturing than $1/\Pi_1$. However, in a more viscous fluid both parameters play a vital role in determining the fracture initiation. The proposed criteria can be used with shear-thinning polymer solutions that are less viscous than Guar presented in this work. In future work, we will examine the fracture initiation and propagation induced by both shear-thinning and shear-thickening fluids. Moreover, more complex granular media including different particle size distribution with multiple layer of particles will be considered.  



\bibliographystyle{plainnat} 
\bibliography{references}  

\end{document}